\documentclass[sigconf]{acmart}
\setcopyright{none}
\renewcommand\footnotetextcopyrightpermission[1]{} 
\usepackage{xspace}
\usepackage{enumitem}
\usepackage{multirow}
\usepackage{colortbl} 
\usepackage{multicol}
\begin{document}

\title{Generative Recommendation with Semantic IDs: \\A Practitioner's Handbook	}

\author{Clark Mingxuan Ju, Liam Collins, Leonardo Neves, \\ Bhuvesh Kumar, Louis Yufeng Wang, Tong Zhao, Neil Shah}
\email{{mju,lcollins2,lneves,bkumar4,ywang14,tong,nshah}@snapchat.com}
\affiliation{%
  \institution{Snap Inc.}
  \city{Santa Monica}
  \state{CA}
  \country{USA}
}

\renewcommand{\shortauthors}{Clark Mingxuan Ju et al.}
\renewcommand{\shorttitle}{Generative Recommendation with Semantic IDs: A Practitioner's Handbook}
\newcommand{\method}{\textsc{GRID}\xspace}
\newcommand{\methodib}{\textsc{GRID}\xspace}

\begin{abstract}
Generative recommendation (GR) has gained increasing attention for its promising performance compared to traditional models. A key factor contributing to the success of GR is the semantic ID (SID), which converts continuous semantic representations (e.g., from large language models) into discrete ID sequences.
This enables GR models with SIDs to both incorporate semantic information and learn collaborative filtering signals, while retaining the benefits of discrete decoding. 
However, varied modeling techniques, hyper-parameters, and experimental setups in existing literature make direct comparisons between GR proposals challenging. 
Furthermore, the absence of an open-source, unified framework hinders systematic benchmarking and extension, slowing model iteration. 
To address this challenge, our work introduces and open-sources a framework for \underline{\textbf{G}}enerative \underline{\textbf{R}}ecommendation with semantic \underline{\textbf{ID}}, namely \method, specifically designed for modularity to facilitate easy component swapping and accelerate idea iteration. 
Using \method, we systematically experiment with and ablate different components of GR models with SIDs on public benchmarks.
Our comprehensive experiments with \method reveal that many overlooked architectural components in GR models with SIDs substantially impact performance. This offers both novel insights and validates the utility of an open-source platform for robust benchmarking and GR research advancement.
\method is open-sourced at \textcolor{blue}{\url{https://github.com/snap-research/GRID}}.
\end{abstract}

\maketitle

\section{Introduction} 
Recommender systems (RecSys) are essential in improving users' experiences on web services, such as product~\citep{schafer1999recommender,ju2025revisiting}, video ~\citep{gomez2015netflix,ju2025learning}, friend~\citep{sankar2021graph,ju2022multi,kolodner2024robust} recommendations.
Among all RecSys, Generative Recommendation (GR) is a rapidly growing paradigm~\citep{rajput2023recommender,zhao2024recommender,lin2025can,zheng2024adapting}, due to recent successes of generative models in vision~\citep{he2022masked,ho2020denoising,dosovitskiy2020image} and language~\citep{guo2025deepseek,brown2020language,bai2023qwen,yu2022generate,ju2022grape}.  
GR leverages advancements in generative models, such as directly generating texts of items of interest~\citep{geng2022recommendation,tan2024idgenrec,hua2023index,bao2023tallrec} or extracting semantic representations from pre-trained models that encode open-world knowledge~\citep{yuan2023go,ren2024representation,yang2024unifying}. 

Within GR, one popular paradigm explores Semantic IDs (SIDs)~\citep{rajput2023recommender,yang2024unifying,deng2025onerec} to bridge the gap between pre-trained foundation models and RecSys. 
As shown in Figure~\ref{fig:1}, this paradigm first harnesses a modality encoder and a quantization tokenizer (e.g., RQ-VAE~\citep{lee2022autoregressive}, VQ-VAE~\citep{esser2021taming}, or Residual K-means~\citep{deng2025onerec}) to transform the modality features (e.g., image or text) into SIDs. A sequential recommender is then trained to autoregressively predict SIDs of future items users will interact with given past items' SIDs.

GR with SIDs provides an effective way to leverage both semantic knowledge encoded in pre-trained foundation models as well as collaborative signals encoding in the user-item interaction history: the overlap in SIDs of two items in principle reflects their semantic similarity, and next-item supervision allows the GR model to learn collaborative signals across SIDs. 
Since the introduction by \cite{rajput2023recommender}, researchers have proposed multiple variants of GR models with SIDs~\citep{deng2025onerec,yang2024unifying,paischerpreference,palumbo2025text2tracks}, several of which advance performance or demonstrate deployment in production systems.

Despite this promising progress, further development of GR with SIDs faces several challenges. 
First, \emph{most relevant literature does not provide open-source implementations}.
This challenge leaves practitioners and researchers to undertake complicated re-implementation burdens, requiring not only advanced technical expertise but also careful hyperparameter tuning. 
This challenge can be further aggravated by the complexity nature of GR pipelines with SIDs.
As we will show later in this paper, good performance of GR pipelines is usually determined by multiple confounding factors at the same time (e.g., appropriate training strategies and careful architecture tuning).
When building a GR pipeline with SIDs from scratch, it is extremely difficult to debug and pinpoint root causes of potential sub-optimal performance, significantly slowing research and development velocity. 
Furthermore, \emph{useful insights around design choices for GR with SIDs are typically not well-discussed in the literature}, leaving practitioners to spend valuable time and computational resources building their own experimental pipelines to develop these understandings on their own.  To bridge these gaps, we make the following contributions:
\begin{figure*}[h]
    \centering
    \includegraphics[width=1\linewidth]{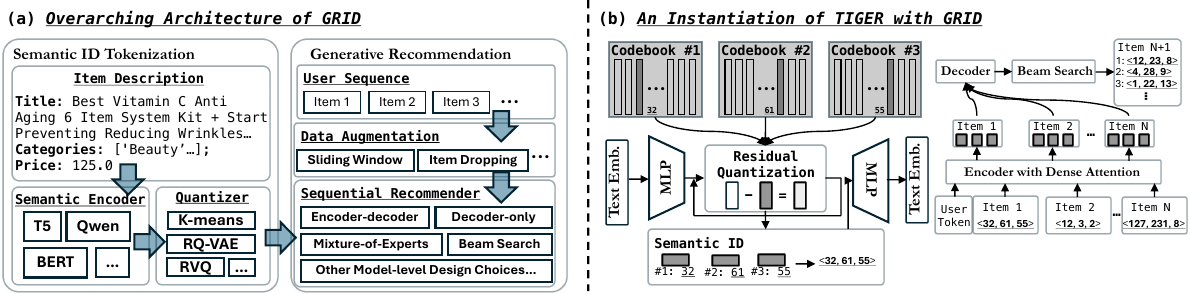}
    \caption{(a) The overarching architecture of GRID. GRID modularizes all intermediate steps in the workflow of GR with SIDs to accelerate the pace of innovation. (b) Instantiating TIGER~\citep{rajput2023recommender} with GRID is straightforward, as it is simply specifying several components already available within GRID, offering practitioners a reliable reference implementation to build on top of. 
    }
    \label{fig:1}
\end{figure*}

\begin{itemize}[leftmargin=*,topsep=0pt]
    \item We present \method, an easy-to-use, flexible, and efficient playground for rapid prototyping of GR with SID methods.  
    \method includes rigorous implementations of fundamental components in GR with SID (e.g., Semantic ID tokenizer, sequential recommender, etc) and can be easily extended to incorporate ongoing efforts in the GR space.  
    To our knowledge, \method is the first open-source resource for prototyping GR with SIDs that can reproduce results reported in existing literature. 
    \item Using \method, we conduct comprehensive experiments to study the impacts of various components in the GR with SID paradigm. 
    Our results include several surprising observations which shed light on key modeling and algorithmic trade-offs that have thus far been mostly overlooked in existing literature.
\end{itemize}

\section{Related Work}

\noindent \textbf{Semantic ID.} RecSys performance hinges on learning quality representations~\citep{cheng2016wide,kim2007music,koren2009matrix,zhao2019recommending,ju2024does,loveland2025understanding}.  Standard practice assigns unique, uninformative IDs to users/items, mapping them to embeddings that capture collaborative signals~\citep{rendle2020neural,yuan2023go,weinberger2009feature,wu2025graphhash,ouyang2025non,loveland2025role}. This approach struggles with scalability~\citep{weinberger2009feature,doublefrequency,Ghaemmaghami2022LearningTC} and performance in sparse or long-tail settings~\citep{shiao2025improving,lika2014facing,vartak2017meta,ju2024does,loveland2025understanding,loveland2025role}. Semantic IDs (SIDs) address these issues by encoding semantic features (e.g., text) via modality encoders (e.g., LLMs), followed by quantization of dense embeddings into sparse IDs~\citep{rajput2023recommender,singh2024better,tan2024idgenrec}. Common quantization-based tokenizers include RQ-VAE~\citep{lee2022autoregressive}, RVQ~\citep{van2017neural}, and Residual K-Means~\citep{deng2025onerec}.

\noindent \textbf{Generative Recommendation with SID.} TIGER~\citep{rajput2023recommender} first applied transformers to predict item SIDs for recommendation, extending ideas from GR in document retrieval~\citep{tay2022transformer}. Follow-up work improves SID training via collaborative signals~\citep{liu2024end,wang2024learnable,zhu2024cost,zheng2025universal,zheng2025pre,xiao2025progressive,chen2024enhancing}, distributional balancing~\citep{kuai2024breaking,yang2025sparse}, and advanced LLMs or multimodal encoders~\citep{liu2024multi,tan2024idgenrec,jin2023language,paischer2024preference,yang2024unifying,zhu2025beyond}. Yet, the field lacks a unified, open-source framework. We address this gap by releasing \method, a modular toolkit to streamline and accelerate GR experimentation.


\section{Proposed Framework: \method -- Generative Recommendation with Semantic IDs}
We consider a set of users $\mathcal{U}$ interacting with items in $\mathcal{I}$. 
Each item $i \in \mathcal{I}$ has associated semantic features $\mathbf{f}_i$, including but not limited to text and image.
Each user $u \in \mathcal{U}$ has an interaction sequence of length $L_u$, denoted as $ \mathcal{S}_u = [i^u_1, i^u_2,\cdots,i^s_{L_u}]$. 
Without loss of generality, a modality encoder (e.g., LLM or VLM) $E(\cdot): \mathbf{f} \rightarrow \mathbb{R}^d$ transforms $\mathbf{f}_i$ into a $d$-dimensional representation $\mathbf{h}_i \in \mathbb{R}^d$.
GR aims to solve the sequential recommendation task, where given a user sequence $\mathcal{S}_u$, GR frameworks generate candidate(s) for the item that the given user will interact with next (i.e., $i^s_{L_u+1}$). 

\subsection{Architecture: Tokenization-then-Generation}
\method splits of GR with SIDs into two separate phases: tokenization-then-generation, adopting the common pattern (see Figure~\ref{fig:1}).
During tokenization, \method maps item embeddings (i.e., $\mathbf{h}_i$) into SIDs. 
During generation, with SIDs for all items, \method explores generative model architectures (e.g., transformer-based models) to generate the SID of $i^s_{L_u+1}$. 
GRID provides flexible implementations for components at each stage.

\noindent \textbf{Semantic ID Tokenization.} SID tokenization entails first computing embeddings $\mathbf{h}_i$ of items' semantic features with a pre-trained modality encoder $E(\cdot)$, then mapping these embeddings to sequences of sparse IDs via a hierarchical clustering tokenizer. 
The hierarchical organization of SID enables precise granularity control via diverse prefix levels.
Formally, given $\mathbf{h}_i$, a tokenizer $\text{Tokenizer}(\cdot): \mathbb{R}^d \rightarrow \{0,1,\cdots,W\}^{L}$ maps item embedding $\mathbf{h}_i$ into a sequence of ID, formulated as 
$\text{SID}_i = \text{Tokenizer}(\mathbf{h}_i)=$ [$\text{SID}^0_i$, $\text{SID}^1_i$, $\cdots$, $\text{SID}^L_i]$, where $W$ refers to the cardinality of each ID and $L$ denotes the number of hierarchies. GRID offers plug-and-play modules for computing $\mathbf{h}_i$, making the swapping of $E(\cdot)$ straightforward -- practitioners can either import customized models or use existing models available on HuggingFace\footnote{\url{https://huggingface.co/docs/transformers}}.
For the tokenzier, GRID supports three algorithms: Residual Mini-Batch K-Means (RK-Means~\citep{deng2025onerec,luo2024qarm}), Residual Vector Quantization (R-VQ~\citep{esser2021taming}), and Residual Quantized Variational Autoencoder (RQ-VAE~\citep{lee2022autoregressive}). 
For instance, as shown in Figure~\ref{fig:1}, the tokenization phase of TIGER~\citep{rajput2023recommender} can be constituted by specifying a T5 encoder to generate item representations, followed by training a RQ-VAE to generate SIDs, all with few-line configuration changes.

\noindent \textbf{Next Item Generation.} With SIDs generated for all items, for each user sequence, GR frameworks leverage a sequential model to generate a list of candidate items that the given user is most likely going to interact with. 
In \method, we incorporate both encoder-decoder~\citep{devlin2019bert,raffel2020exploring,vaswani2017attention} and decoder-only~\citep{brown2020language,bai2023qwen,touvron2023llama} model architectures with flexible configurations (e.g., the number of heads, layers, or mixture-of-experts in transformer layers). As for tokenization, practitioners can easily import custom architectures or borrow model architectures publicly available in HuggingFace. 
By default, we train the generation model using the commonly adopted next-token prediction objective~\citep{rajput2023recommender,kang2018self} with sliding window augmentation~\citep{zhou2024contrastive}\footnote{Training and inference logic can be fully customized.}.
The inference generation is conducted through beam search with KV-cache, with tunable hyper-parameters such as beam width, whether search is restricted to valid SIDs, etc. 
We also provide implementations for several tricks broadly explored in existing literature to showcase {\method}'s flexibility, including user token~\citep{rajput2023recommender}, and de-duplication of SID to avoid collision.
\begin{table*}[h]
\vspace{-0.1in}
\caption{
Performance of GR models with by SIDs generated by different tokenization algorithms.
}
\vspace{-0.15in}
\resizebox{1.91\columnwidth}{!}{
    \begin{tabular}{l|cccc|cccc|cccc}
    \toprule
    \multirow{2}{*}{Methods} & \multicolumn{12}{c}{Recommendation Performance (Recall@5/Recall@10/NDCG@5/NDCG@10)} \\
    \cmidrule(r){2-13}
    &\multicolumn{4}{c}{Beauty}&\multicolumn{4}{c}{Toys}&\multicolumn{4}{c}{Sports}\\
    \midrule
    RK-Means & \textbf{0.0422} & \textbf{0.0639}  & 0.0277  & 0.0347
    & \textbf{0.0376} & \textbf{0.0577}  & \textbf{0.0243}  & \textbf{0.0308} 
    & \textbf{0.0236} & \textbf{0.0353}  & \textbf{0.0153} & \textbf{0.0191}  \\
    R-VQ & \textbf{0.0422}  & 0.0638  & \textbf{0.0282}  & \textbf{0.0351} 
    & 0.0327 & 0.0493  & 0.0209  & 0.0262 
    & 0.0234 & 0.0352  & 0.0151  & 0.0189 \\
    RQ-VAE & 0.0404 & 0.0593 & 0.0268  & 0.0329
    & 0.0342 & 0.0514 & 0.0224  & 0.0280
    & 0.0205 & 0.0312 & 0.0132 & 0.0166  \\
    \bottomrule
    \end{tabular}
    }
\label{tab:rq_model_type}
\end{table*}

\begin{table*}[h]
\vspace{-0.08in}
\caption{Performance of GR models with SIDs generated by RK-Means with varying-size language model encoders.
}
\vspace{-0.15in}
\resizebox{1.89\columnwidth}{!}{
    \begin{tabular}{l|cccc|cccc|cccc}
    \toprule
    \multirow{2}{*}{LM} & \multicolumn{12}{c}{Recommendation Performance (Recall@5/Recall@10/NDCG@5/NDCG@10)} \\
    \cmidrule(r){2-13}
     &\multicolumn{4}{c}{Beauty}&\multicolumn{4}{c}{Toys}&\multicolumn{4}{c}{Sports}\\
    \midrule
        L & \textbf{0.0429} & 0.0639 & \textbf{0.0285} & \textbf{0.0353}
    & 0.0373 & 0.0565 & 0.0237 & 0.0300
    & 0.0224 & 0.0347 & 0.0145 & 0.0185 \\
    XL & 0.0422 & 0.0639 & 0.0277 & 0.0347 
    & 0.0376 & 0.0577 & 0.0243  & 0.0308 
    & 0.0236 & 0.0353 & 0.0153  & 0.0191 \\
    XXL & \textbf{0.0429} & \textbf{0.0646} & 0.0282  & 0.0352 
    & \textbf{0.0381} & \textbf{0.0586} & \textbf{0.0245}  & \textbf{0.0311}
    & \textbf{0.0239} & \textbf{0.0363} & \textbf{0.0154}  & \textbf{0.0194}  \\
    \bottomrule
    \end{tabular}
    }
\label{tab:lm_dimension}
\end{table*}

\begin{table}[h]
\vspace{-0.1in}
\caption{
Beauty GR performance with RK-Means SIDs with varying codebook dimensions from Flan-T5-XL embeddings.
}
\vspace{-0.15in}
\resizebox{0.95\columnwidth}{!}{
    \begin{tabular}{l|cccc}
    \toprule
    $L\times W$ & Recall@5 & Recall@10 & NDCG@5 & NDCG@10\\
    \midrule
    3$\times$128 & 0.0412 & 0.0617 & 0.0273 & 0.0339\\
    3$\times$256 & \textbf{0.0422} & \textbf{0.0639} & \textbf{0.0277} & \textbf{0.0347}\\
    3$\times$512 & 0.0415 & 0.0631 & 0.0273 & 0.0342\\
    2$\times$256 & 0.0403 & 0.0618 & 0.0264 & 0.0333 \\
    4$\times$256 & 0.0405 & 0.0609 & 0.0265 & 0.0331\\
    5$\times$256 & 0.0396 & 0.0596 & 0.0257 & 0.0321\\
    \bottomrule
    \end{tabular}
    }
\label{tab:sid_dimension}
\end{table}


\section{Experiments with \method}

We next showcase GRID's utility by conducting a brief, but rigorous, investigation into the performance trade-offs around several fundamental, yet overlooked, design choices within GR with SID.

\noindent \textbf{Setup.} We evaluate on 5-core filtered Amazon Beauty, Sports, and Toys datasets~\citep{hou2024bridging,rajput2023recommender}, using the last item per user for test, second-to-last for validation, and the rest for training. Item text features include Title, Categories, Description, and Price. Semantic embeddings are extracted via mean pooling over final hidden states of Flan-T5-Large, XL, and XXL~\citep{chung2024scaling}.  For tokenization, we consider RK-Means~\citep{deng2025onerec}, R-VQ~\citep{esser2021taming}, and RQ-VAE~\citep{lee2022autoregressive}. For generation, we analyze architectural choices mentioned above. Tokenizers are trained on 8 GPUs with per-device batch size 2048. RK-Means and R-VQ are trained layer-wise for 1k steps per layer; RQ-VAE for 15k total steps. The learning rate (LR) is $10^{-3}$ with Adam (R-VQ) or Adagrad (RQ-VAE). Residuals are normalized (RK-Means, R-VQ) and embeddings whitened (RQ-VAE) to prevent collapse. Generative models use Adam with LR $5\times10^{-4}$, weight decay $10^{-6}$, and batch size 256. We use sliding-window sampling~\citep{paischer2024preference}, with early stopping after 10 validation intervals (100 steps each) without NDCG@10 improvement. 
For model architectures of generative models, we explore 8 transformer layers in total (i.e., 4 in the encoder and 4 in the decoder for encoder-decoder models) with 6 attention heads in each layer, an embedding dimension of 128, and 1024 hidden dimensions for MLP layers.
We report Recall@K and NDCG@K for $K\in{5,10}$ on the test set, using the checkpoint with best validation Recall@10. All results are averaged over 5 runs with different seeds.

\subsection{Semantic ID Tokenization}
We study SID tokenization by ablating (1) the choice of SID tokenizer algorithm, (2) the size of the pre-trained semantic encoder, and (3) the number of residual layers (i.e., $L$) and tokens per layer (i.e., $W$) in the SID tokenizer.
We train variants of SID tokenizers and evaluate the performance of the base sequential recommendation model trained using the corresponding tokenizer. Unless otherwise noted, we use RK-Means with $(L,W)=(3,256)$ and Flan-T5-XL.

\noindent\textbf{SID Tokenizer Algorithm.}
RQ-VAE is commonly adopted in the literature as the default SID tokenizer
\citep{zheng2024adapting,paischerpreference,yang2025sparse,singh2024better} since its use by TIGER \citep{rajput2023recommender}. However, it entails simultaneously training an autoencoder and quantizer, exacerbating a number of challenges \citep{kuai2024breaking,fifty2024restructuring,zhu2024addressing} and raising the question of whether its performance benefits are worth the implementation complexity. Table \ref{tab:rq_model_type} suggests the answer is ``no'': RK-Means and sometimes R-VQ lead to better recommendation performance than RQ-VAE, despite our training of RQ-VAE for 5x as many iterations as the simpler alternatives.

\noindent\textbf{Semantic Encoder Size.} We next vary the size of the Flan-T5 model \citep{chung2024scaling}  used to compute the semantic embeddings,  from Large (780M parameters) to XL (3B) to XXL (11B). Table \ref{tab:lm_dimension} shows only marginal increases in recommendation performance due to  an over 14-fold increase in the number of LLM parameters, indicating that the current GR with SID pipeline can be improved by more fully leveraging the increased world knowledge in larger LLMs.

\noindent\textbf{SID Tokenizer Dimension.}
In Table \ref{tab:sid_dimension}, we vary the number of residual layers $L$ and tokens per layer $W$ in RK-Means, and observe that the default choice of $(L,W)=(3,256)$ leads to the best recommendation performance. Surprisingly, performance drops substantially with more layers, although additional layers convey more semantic information to the recommendation model. This points to a trade-off between SID sequence learnability and the amount of semantic information contained in the SIDs.

\begin{table}
\vspace{-0.1in}
\caption{
\centering
GR performance on Beauty with varying number of TIGER~\citep{rajput2023recommender}-style user tokens.
}
\vspace{-0.15in}
\resizebox{0.95\columnwidth}{!}{
    \begin{tabular}{l|cccc}
    \toprule
    \# Tokens & Recall@5 & Recall@10 & NDCG@5 & NDCG@10\\
    \midrule
    0 & \textbf{0.0408} & \textbf{0.0618} & \textbf{0.0270} & 0.0330\\
    2,000 & 0.0396 & 0.0597 & 0.0264 & 0.0328\\
    4,000 & 0.0401 & 0.0612 & 0.0264 & 0.0332\\
    6,000 & 0.0401 & 0.0611 & 0.0264 & 0.0331\\
    8,000 & 0.0405 & 0.0610 & 0.0269 & \textbf{0.0335}\\
    \bottomrule
    \end{tabular}
    }
\label{tab:user_token}
\end{table}
\vspace{-2mm}
\begin{table}
\caption{
GR performance with different architectures.
}
\vspace{-0.15in}
\resizebox{0.95\columnwidth}{!}{
    \begin{tabular}{l|cccc}
    \toprule
    Model & Recall@5 & Recall@10 & NDCG@5 & NDCG@10\\
    \midrule
    \multicolumn{5}{c}{Beauty} \\
    \midrule
    Enc-Dec & \textbf{0.0396} & \textbf{0.0597} & \textbf{0.0264} & \textbf{0.0328} \\
    Dec-only & 0.0300 & 0.0438 & 0.0206 & 0.0251 \\
    \midrule
    \multicolumn{5}{c}{Toys} \\
    \midrule
    Enc-Dec & \textbf{0.0357} & \textbf{0.0548} & \textbf{0.0226} & \textbf{0.0287} \\
    Dec-only & 0.0286 & 0.0399 & 0.0202 & 0.0238 \\
    \midrule
    \multicolumn{5}{c}{Sports} \\
    \midrule
    Enc-Dec & \textbf{0.0192} & \textbf{0.0290} & \textbf{0.0124} & \textbf{0.0156} \\
    Dec-only & 0.0152 & 0.0226 & 0.00979 & 0.0121 \\
    \bottomrule
    \end{tabular}
    }
\label{tab:encdec}
\end{table}
\vspace{-1mm}
\begin{table}
\caption{
GR performance with,  without data augmentation.
}
\vspace{-0.15in}
\resizebox{0.95\columnwidth}{!}{
    \begin{tabular}{l|cccc}
    \toprule
    Augmentation & Recall@5 & Recall@10 & NDCG@5 & NDCG@10\\
    \midrule
    \multicolumn{5}{c}{Beauty} \\
    \midrule
    Sliding Window & \textbf{0.0396} & \textbf{0.0597} & \textbf{0.0264} & \textbf{0.0328} \\
    No Augmentation & 0.0279 & 0.0447 & 0.0171 & 0.0226 \\
    \midrule
    \multicolumn{5}{c}{Toys} \\
    \midrule
    Sliding Window & \textbf{0.0357} & \textbf{0.0548} & \textbf{0.0226} & \textbf{0.0287} \\
    No Augmentation & 0.0277 & 0.0442 & 0.0173 & 0.0226 \\
    \midrule
    \multicolumn{5}{c}{Sports} \\
    \midrule
    Sliding Window & \textbf{0.0192} & \textbf{0.0290} & \textbf{0.0124} & \textbf{0.0156} \\
    No Augmentation & 0.0174 & 0.0250 & 0.0114 & 0.0140 \\
    \bottomrule
    \end{tabular}
    }
\label{tab:aug}
\end{table}

\begin{table}
\caption{
GR performance with, without SID de-duplication.
}
\vspace{-0.15in}
\resizebox{0.95\columnwidth}{!}{
    \begin{tabular}{l|cccc}
    \toprule
    Approach & Recall@5 & Recall@10 & NDCG@5 & NDCG@10\\
    \midrule
    \multicolumn{5}{c}{Beauty} \\
    \midrule
    With De-dup. & \textbf{0.0396} & \textbf{0.0597} & \textbf{0.0264} & \textbf{0.0328} \\
    No De-dup. & 0.0381 & 0.0591 & 0.0253 & 0.0321 \\
    \midrule
    \multicolumn{5}{c}{Toys} \\
    \midrule
    With De-dup. & \textbf{0.0357} & \textbf{0.0548} & \textbf{0.0226} & \textbf{0.0287} \\
    No De-dup. & 0.0353 & 0.0532 & 0.0225 & 0.0282 \\
    \midrule
    \multicolumn{5}{c}{Sports} \\
    \midrule
    With De-dup. & \textbf{0.0192} & \textbf{0.0290} & \textbf{0.0124} & \textbf{0.0156} \\
    No De-dup. & 0.0186 & 0.0269 & 0.0011 & 0.0142 \\
    \bottomrule
    \end{tabular}
    }
\label{tab:dedup}
\end{table}

\begin{table}
\caption{
GR performance with different beam searches.
}
\vspace{-0.15in}
\resizebox{0.95\columnwidth}{!}{
    \begin{tabular}{l|cccc}
    \toprule
    Approach & Recall@5 & Recall@10 & NDCG@5 & NDCG@10\\
    \midrule
    \multicolumn{5}{c}{Beauty} \\
    \midrule
    Constrained & 0.0396 & 0.0597 & 0.0264 & 0.0328 \\
    Free-form & \textbf{0.0405} & \textbf{0.0609} & \textbf{0.0268} & \textbf{0.0334} \\
    \midrule
    \multicolumn{5}{c}{Toys} \\
    \midrule
    Constrained & \textbf{0.0357} & \textbf{0.0548} & 0.0226 & 0.0287 \\
    Free-form & 0.0356 & 0.0546 & \textbf{0.0227} & \textbf{0.0289} \\
    \midrule
    \multicolumn{5}{c}{Sports} \\
    \midrule
    Constrained & 0.0192 & 0.0290 & 0.0124 & 0.0156 \\
    Free-form & \textbf{0.0198} & \textbf{0.0302} & \textbf{0.0127} & \textbf{0.0160} \\
    \bottomrule
    \end{tabular}
    }
\label{tab:beam}
\end{table}
\subsection{Generative Recommendation}
To understand the impact of next-item generation model design choices, we ablate the: (1) quantity of user tokens, (2) choice between encoder-decoder and decoder-only architectures, (3) integration of data augmentation for training, (4)  implementation of de-duplication IDs, and (5) employment of constrained or unconstrained beam search.
We employ the default SID tokenization described previously and evaluate next-item generation performance.

\noindent\textbf{The Quantity of User Tokens.} TIGER~\citep{rajput2023recommender} prepends a user token to every user's SID sequence, where user tokens are assigned via a random hash into a fixed vocabulary size. 
Table~\ref{tab:user_token} reveals that a larger user token vocabulary does not always improve performance and totally removing this design (i,.e., 0) leads to the optimal performance, implying that the current standard use of user tokens in GR with SIDs is not achieving its goal of personalization.

\noindent\textbf{Encoder-decoder vs. Decoder-only Architecture.}
Most existing literature explores encoder-decoder based transformer generative models~\citep{yang2024unifying,rajput2023recommender}. To investigate the feasibility of decoder-only architecture, we swap the encoder-decoder backbone with a decoder-only architecture. 
As shown in Table~\ref{tab:encdec}, decoder-only models significantly under-perform encoder-decoder models. 
We hypothesize that this substantial performance gap can be attributed to the inherent design of encoder-decoder models, where the encoder's dense attention mechanism over the entire user history effectively captures richer and more comprehensive sequential patterns. This deep contextual understanding, subsequently leveraged by the decoder for generation, appears to be critical for the challenging task of generative recommendation. 

\noindent\textbf{Data Augmentation for Training.}
In Table~\ref{tab:aug}, we study the impact of data augmentation on the performance of GR with SIDs. We explore sliding window data augmentation, where a single user sequence is expanded into all possible contiguous sub-sequences~\citep{zhou2024contrastive}. 
Our observations strongly indicate that proper data augmentation is paramount for achieving robust and high-performing GR models.
The expanded and varied training samples generated through this technique likely enhance the model's ability to learn more generalizable patterns from user interactions, mitigate overfitting, and improve its capacity to predict diverse next items, even in the presence of noisy and/or sparse data. 

\noindent\textbf{De-duplication of SIDs.}
De-duplicating SIDs is essential for accurate retrieval. We compare two strategies: TIGER's method, which appends a digit to SIDs to resolve collisions ("With De-dup."), and a simpler approach of randomly selecting an item when SIDs conflict. Table \ref{tab:dedup} shows both perform comparably, with TIGER's strategy having a slight edge. However, TIGER's approach increases sequence length and decoding complexity, and its requirement for global SID distribution knowledge is impractical for large item sets.

\noindent\textbf{Constrained vs. Unconstrained Beam Search.}
Decoding strategy impacts both generation quality and computational efficiency. Our ablation study compares constrained and unconstrained beam search, vital for generative recommender system deployment. While constrained beam search guides output to valid SIDs, unconstrained explores all sequences without explicit rules. Table~\ref{tab:beam} shows both yield similar performance. Crucially, unconstrained beam search was significantly more efficient and computationally cheaper. This suggests the SID generation task's inherent structure, combined with learned model patterns, is sufficient for high-quality recommendations without the overhead of explicit constraints.

\section{Conclusion}

In this work, we highlight the critical need for a unified open-source framework in GR with SIDs. 
Through \method, we conduct systematic experiments, revealing multiple surprising insights. We discover that several components previously assumed essential—and which are often compute- and/or engineering-intensive—can actually be replaced with more efficient alternatives without sacrificing performance. Conversely, other, often overlooked design choices prove to be vital, such as encoder-decoder architectures and data augmentation. These findings not only offer novel insights into the true drivers of GR with SID performance but also underscore the immense value of an open-source platform like \method for robust benchmarking and accelerating research.
\bibliographystyle{ACM-Reference-Format}
\bibliography{sample-base}

\end{document}